# Influence of interlayer coupling on the spin torque driven excitations in a spin torque oscillator

M. Romera[1,2,3], E. Monteblanco[1,2,3], F. Garcia-Sanchez[1,2,3], B. Delaët[4],

L. D. Buda-Prejbeanu[1,2,3], U. Ebels[1,2,3]

[1]Univ. Grenoble Alpes, F-38000 Grenoble, France
[2]CEA, INAC-SPINTEC, F-38000 Grenoble, France
[3]CNRS, SPINTEC, F-38000 Grenoble, France
[4]CEA-LETI, MINATEC, DRT/LETI/DIHS, 38054 Grenoble, France

The influence of dynamic interlayer interactions on the spin torque driven and damped excitations are illustrated for a three layer macrospin model system that corresponds to a standard spin-torque oscillator. The free layer and a synthetic antiferromagnetic (SyF) pinned layer of the spin-torque oscillator are in-plane magnetized. In order to understand experimental results, numerical simulations have been performed considering three types of interlayer interactions: exchange interaction between the two magnetic layers of the SyF, mutual spin torque between the top layer of the SyF and the free layer and dipolar interaction between all three magnetic layers. It will be shown that the dynamic dipolar coupling plays a predominant role. First, it leads to a hybridization of the free layer and the SyF linear modes and through this gives rise to a strong field dependence of the critical current. In particular, there is a field range of enhanced damping in which much higher current is required to drive the modes into steady state. This results in a gap in the excitation spectrum. Second, the dynamic dipolar interaction is also responsible for the non-linear interaction between the current driven steady state mode and the damped modes of the system. Here one can distinguish: (i) a resonant interaction that leads to a kink in the frequency-field and frequency-current dispersions accompanied by a small hysteresis and a reduction of the linewidth of the steady state mode and (ii) a non-resonant interaction that leads to a strong frequency redshift of the damped mode. The results underline the strong impact of interlayer coupling on the excitation spectra of spin-torque oscillators and illustrate in a simple three mode model system how in the non-linear regime the steady state and damped modes influence each other.

# I. INTRODUCTION

Spin-torque oscillators (STOs) are promising candidates for integrated radiofrequency devices [1,2,3]. The optimization of their microwave properties such as frequency range, frequency dispersion vs current and field, output power and linewidth has led to an intensive study of different STO structures over the last few years. The most standard STO contains an in-plane magnetized single free layer (FL) and a synthetic ferrimagnetic (SyF) layer as a polarizer, which consists of two in-plane magnetized ferromagnetic layers coupled antiferromagnetically through a thin non-magnetic spacer via interlayer exchange coupling (RKKY interaction). In the first theoretical[4,5] and experimental studies the free layer was considered as a single non-interacting system which, under application of a spin polarized current, behaves dynamically as an independent entity while the magnetization of the SyF remained fixed. This picture provides only a general understanding of the most basic features of spin torque excitations in magnetic nanopillars. However, important questions remain open and many features observed experimentally cannot be explained in the frame of this 'independent free layer' picture. Recent investigations have addressed the coupling between the different magnetic layers of an STO and demonstrated that it strongly impacts the excitations spectra[6-18], and that it can be exploited in the interest of the device performances[7-10].

In order to understand the role of the different coupling mechanisms within a standard STO structure, one has to consider one by one the different interactions for simple model systems. This has been done in previous work [6-10, 14-15].

One of the first studies going beyond a non-interacting independent free layer, addressed via numerical macrospin simulations the spin torque driven excitations of a two layer RKKY coupled system in the form of a SyF excited through an external polarizer (ref. [6]). As a specific characteristics of this coupled SyF structure it was found that the frequency $f$ - current $I_{app}$ dependence of the steady state oscillations changes from frequency redshift ($df/dI_{app}<0$) to frequency blueshift ($df/dI_{app}>0$) upon increasing the in-plane applied field. This has been confirmed experimentally on spin valve structures [11]. This behaviour is in contrast to the independent free layer which is characterized by only a frequency redshift. The numerical analysis has been extended to show that the RKKY coupling strength in conjunction with an asymmetry of internal fields of the two layers is responsible for this change from redshift to blueshift (ref. [7]). For symmetric structures and weak RKKY interactions, one recovers the behaviour of the independent free layer.

In a next step the external polarizer was removed from the SyF structure to consider a simple system of two layers coupled by conservative (RKKY) as well as dissipative (mutual spin torque) interactions. This self-polarized structure was studied numerically [8] and showed a similar change from redshift to blueshift. In an analytical study [14], the spin wave formalism of single layers [5] was extended to the self-polarized coupled structure. It was demonstrated that besides conservative terms of the hamiltonian also the dissipative terms need to be considered to explain the redshift to blueshift transition. This formalism provides a theoretical framework for the resonant and non-resonant interactions within the Hamiltonian that is of relevance for the results presented here.

In a first approach to a standard STO, a three layer system was considered in [6, 9] via numerical macrospin simulations by coupling the free layer to the SyF via dissipative interaction (mutual spin torque). It was shown that this non-conservative coupling can lead to a non-hysteretic resonant interaction between the free layer and the SyF when their respective frequencies are close. This translates to a strong deviation in the frequency of the spin torque driven free layer mode that locks to the frequency of the damped mode of the SyF. It was furthermore shown that this frequency locking is accompanied by a reduction of the linewidth via the increase of the amplitude relaxation rate and the reduction of the non-linear amplitude-phase coupling parameter ν [9].

A recent work [10] added to this dissipative coupling between the FL and the SyF of the standard STO the dipolar interaction between all three layers. It was shown that this dipolar interaction can also give rise to a linewidth reduction through a non-linear interaction between the FL dominated spin torque driven mode with one of the SyF dominated damped modes. In the present manuscript we provide a more detailed analysis of this *non-linear* interaction and argue that in view of the analytical model of ref. 14 it is a *resonant* interaction between the spin torque driven and the damped mode. We furthermore discuss in detail the role of the dipolar interaction on the linear modes of the coupled system and show that this coupling affects the transition of these modes into steady state oscillations leading to gaps in the excitation spectra. Finally we show the impact of *non-linear non-resonant* interactions on the frequencies of damped modes. For this a systematic macrospin analysis of the linear and non-linear excitations is carried out. The numerical results are compared to the experimental results.

The manuscript is organized as follows: the device structure used experimentally and in the numerical simulations are presented in section II together with the simulation parameters used and the interactions considered. Section III presents the analysis of the damped modes of the coupled three layer system to derive the state diagram and the critical boundaries. The non-linear steady state regime and the associated non-linear effects are discussed in section IV. The results are summarized in section V.

## II. TECHNIQUES

### A. Experiments

Experimental results were obtained from spin valve (SV) nanopillar devices of the following composition: PtMn(20)/SyF/Cu(3)/FL, where SyF corresponds to the polarizer CoFe(2.5)/Ru(0.8)/[CoFe(1)/Cu(0.3)]$_2$/CoFe(1), FL is the free layer CoFe(1)/Ni$_{80}$Fe$_{20}$(2) and numbers represent thickness in nanometers[10]. The SV were grown by sputter-deposition using a SINGULUS TIMARIS tool. Using a combination of electron beam lithography and ion milling, the films were then patterned into elliptical pillars with axis dimensions of 140 x 70 nm. Figure 1a shows a typical MR loop of these devices at low current. Positive current is defined when the electrons flow from the SyF to the free layer. Therefore, a positive current destabilizes the FL in the antiparallel state (AP), which is the state considered here.

### B. Simulations

In the simulations a standard STO structure similar to the experimental one is considered, composed by an in-plane magnetized free layer and an in-plane magnetized SyF polarizer (see schematic inset in Fig. 1a). The SyF consists of two in-plane magnetized ferromagnetic layers [bottom layer (BL) and top layer (TL)] exchange coupled through a non-magnetic Ru spacer. The SyF is pinned by an antiferromagnet.

The magnetization dynamics of this system is described by solving in a macrospin approximation the Landau-Lifshitz-Gilbert (LLG) equation

$$\frac{d\boldsymbol{M}_i}{dt} = -\gamma_0(\boldsymbol{M}_i \times \boldsymbol{H}_i^{eff}) + \frac{\alpha}{M_{Si}}\left(\boldsymbol{M}_i \times \frac{d\boldsymbol{M}_i}{dt}\right) + \gamma_0 \frac{a_J}{M_{Si}} \boldsymbol{M}_i \times (\boldsymbol{M}_i \times \boldsymbol{M}_j)$$

(1)

simultaneously for all three layers i,j= FL,TL,BL. Here $\gamma_0 = \mu_0\gamma$ is the gyromagnetic ratio of the free electron multiplied by the vacuum magnetic permeability, $a_J$ is a factor proportional to the applied current, $M_{Si}$ is the spontaneous magnetization of the *i*th layer, $\boldsymbol{M}_i$ the corresponding magnetization vector and $\boldsymbol{H}_i^{eff} = \boldsymbol{H}_i^{int} + \boldsymbol{H}_i^{coupling}$ is the effective field, where $\boldsymbol{H}_i^{int}$ is the intrinsic field of the *i*th layer (sum of anisotropy, demagnetizing fields, external field and exchange bias for the BL) and $\boldsymbol{H}_i^{coupling}$ is the coupling field of the *i*th layer with other magnetic layers.

In the simulations, three types of dynamic couplings are taken into account: dynamic RKKY interaction between the two magnetic layers of the SyF (BL and TL), dynamic dipolar interaction between the three magnetic layers of the STO and mutual spin torque (MSTT) between the FL and the TL of the SyF. Dynamic RKKY interaction and dynamic dipolar interaction are conservative couplings included in the precession term of the LLG equation (first term in Eq. 1). MSTT is a dissipative coupling taken into account in the spin torque term (last term in Eq. 1). To simulate the effect of finite temperature T=400K, a fluctuating thermal noise field is added in Eq. 1. The parameters of the magnetic layers used in the simulations are listed in table 1. The thickness of the Ru and Cu spacers and the dipolar coefficients have been chosen to mimic the experimental samples. Under these conditions, the strength of the static dipolar field at the different layers in the zero applied field is 125 Oe for the FL, -641 Oe for the TL and -7 Oe for the BL, while the intrinsic fields are -597 Oe, -1715 Oe and 2193 Oe respectively. The RKKY interlayer exchange energy is taken as J$_{RKKY}$=-1mJ/m².

|  | BL | TL | FL |
|---|---|---|---|
| $M_S$ (kA/m) | 1600 | 1340 | 1070 |
| $K_u$ (J/m$^3$) | 8000 | 6700 | 5350 |
| $t$ (nm) | 2.5 | 3 | 3 |
| $\alpha$ | 0.02 | 0.02 | 0.02 |
| $\eta$ | - | 0.3 | 0.3 |
| $H_{ex}$ (Oe) | 500 | 0 | 0 |

Table 1: Parameters used in the numerical simulations. $H_{ex}$ is the exchange bias field of the SyF bottom layer pinned by an antiferromagnet.

## III. CRITICAL BOUNDARIES AND LINEAR EIGENMODES

In this section we first present experimental results on the state diagram that are then analyzed numerically by calculating the linear eigenmodes in the subcritical regime of the coupled three layer system. From this the influence of the interlayer coupling on the boundary of the critical current vs applied field is established. The influence of the interlayer coupling on the linear and steady state modes beyond the critical current will be analyzed in section IV.

### III.A. Experiments

Fig. 1b shows the experimental state diagram that summarizes the static (blue color) and dynamic regions (red color) within the AP state as a function of field and current. The most striking feature of the experimental state diagram is the strong field dependence of the critical current both for positive and negative current polarities. As a consequence, for certain values of applied current the steady state excitations vanish in a particular region of field, which translates into 'gaps' in the frequency-field dispersion curve (see Fig. 1c) of the steady state mode. These gaps, which are highlighted in the state diagram by white arrows (Fig.1b), are observed both for positive and negative current polarities, and in both cases they are reduced upon increasing current (Fig. 1d). The gap of $j_{app}<0$ is more pronounced and is not closed within the current range used in the experiment. This range was limited by heating or destruction of the devices. This behavior of the critical current dependence contrasts the one expected in the 'independent free layer picture', where the critical current of instability follows a continuous line and depends only weakly on the field[6].

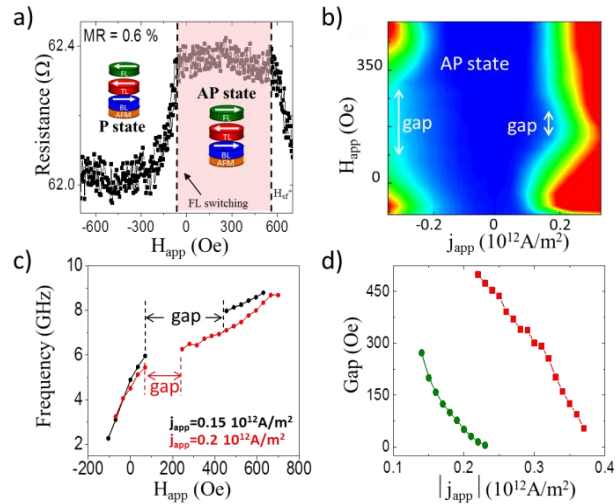

Figure 1: (a) Magnetoresistance loop of the device measured at low current. (b) Experimental state diagram where blue color represents static antiparallel state and red color represents dynamic excitations. (c) Frequency field dispersions at $j_{app}=0.15 \cdot 10^{12}$ A/m$^2$ (black curve) and $j_{app}=0.22 \cdot 10^{12}$ A/m$^2$ (red curve),

obtained for an elliptical pillar with axis dimensions of 140 x 70 nm at room temperature. (d) Range of field where excitations vanish (gap size) as a function of the applied current for positive (green symbols) and negative (red symbols) current polarities.

### III.B. Stability analysis

### III.B.1. Linear eigen-mode frequencies for $j_{app}=0$

In order to understand the strong dependence of the critical current on the applied field and the regions of absence of steady state, we have analyzed numerically the linear modes of the coupled system and how these modes are driven into steady state. The first step is to determine the three linear eigenmodes of the system in absence of current. For this the coupled LLG equations of the three layers have been linearized around the AP state and then their characteristic frequencies were determined numerically. Linear modes follow the relation $\boldsymbol{M}(t) \sim \boldsymbol{M_0} + \Delta \boldsymbol{M} e^{2\pi(\lambda+if)t}$ where the real part $\lambda$ gives the attenuation and the imaginary part $f$ the frequency. The resulting frequency-field dependencies of the three $j_{app}=0$ eigenmodes are shown in Fig. 2a-top. In absence of dipolar interlayer coupling, the frequency-field dispersion at $j_{app}=0$ for a single non-interacting free layer and for the acoustic (ac) and optic (op) mode of the SyF are obtained (black and blue dashed lines respectively in Figure 2a-top). It can be seen that there is a crossing of the FL and ac-SyF mode frequencies. When dipolar interlayer interaction is taken into account the FL and ac-SyF eigenmodes split into a binding and antibinding mode (modes 1 and 2 red, green symbols respectively in Fig.2a-top), while the optic mode of the SyF is slightly shifted upwards in frequency resulting in mode 3 (violet). In all three coupled modes all three magnetic layers oscillate simultaneously.

### III.B.2. Attenuation of the eigenmodes

Let us focus now on the attenuation $\lambda$ of the three coupled modes in absence of applied current (Fig. 2a-bottom, red and green and violet lines for modes 1, 2 and 3 respectively). In absence of dipolar interlayer interaction the dashed lines in Fig. 2a-bottom represent the attenuation of the single non-interacting free layer mode (black) and ac-SyF mode (blue). It is noted that the absolute value of the attenuation increases for the FL and decreases for the ac-SyF mode upon increasing field. In presence of the dipolar interlayer interaction the full lines represent the attenuation of the coupled modes 1, 2 and 3. Let us first consider the attenuation of mode 1. As can be seen in Fig. 2a, both the frequency and the attenuation of mode 1 correspond mainly to the frequency and attenuation of the independent FL at low fields and to those of the independent ac-SyF at high fields. On the other hand, in the field region around the splitting the values of frequency and attenuation of mode 1 lie in between those of the independent layers. This behavior suggests that mode 1 is strongly dominated by the FL at low fields and by the SyF at high fields, while in the intermediate field region around the splitting it is a strongly hybridized mode to which FL and SyF contribute almost equally. To corroborate this assumption Fig. 2d shows the FFT versus frequency of the simulated $M_y$ component of the FL (Fig. 2d-left column) and the BL of the SyF (Fig. 2d-right column) for $j_{app}=0$ and three different field values, where $M_y$ is the in-plane component of the magnetization in the direction perpendicular to the applied field.

At zero field, the $M_y$ component of the FL (Fig. 2d-1) shows only one strong peak corresponding to mode 1 while the $M_y$ component of the BL (Fig. 2d-4) gives a very weak contribution at the frequency of mode 1. In contrast, at high field the strongest contribution to mode 1 is observed in the BL (Fig. 2d-6) while the FL contribution is very weak (Fig. 2d-3). On the other hand in the field region around the splitting two peaks of similar intensities can be seen both in the $M_y$ component of the FL (Fig. 2d-2) as in the BL (Fig. 2d-5). This confirms our assumption of mode 1 being FL-dominated at low fields, SyF-dominated at high fields, and highly hybridized (with equal contribution from the FL and the SyF) at intermediate fields. Similarly, mode 2 is dominated by the SyF at low fields and by the FL at high fields while contributions are similar around the splitting. It is noted that there is a considerable contribution of the SyF to mode 3 in the field range considered, while the contribution of the FL to mode 3 remains weak.

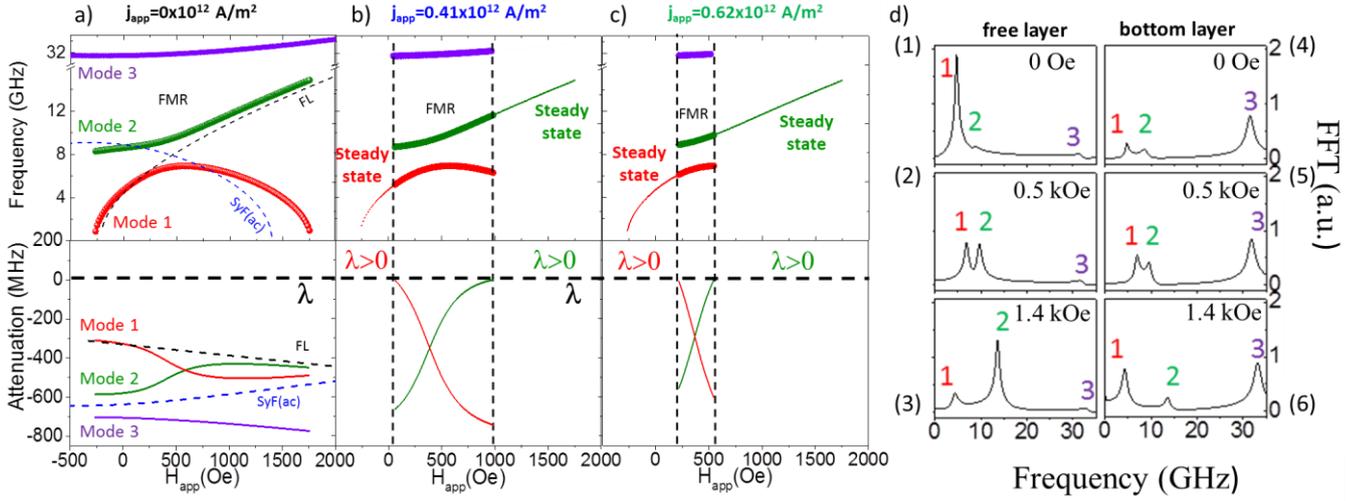

Figure 2: (a-c,top) Frequency-field dependence of the three linear modes of the coupled system (red, green and violet symbols respectively) at different positive currents. Full symbols represent frequencies of the linear eigenmodes (i.e. negative attenuation) while continuous lines in b and c are guides to the eye to highlight the modes that are in steady state (note that the real frequency of the steady state mode will be redshifted). Dashed black and blue lines in a represent the frequency of the free layer and the SyF in the 'uncoupled' (independent) picture. (a-c, bottom) Field dependence of the attenuation of the three linear modes. Dashed black and blue lines in a represent attenuation of a single non-interacting free layer and a SyF in the 'uncoupled' (independent) picture. (d) FFT of the simulated $M_y$ component of the FL (1-3) and SyF (4-6) upon increasing field for $j_{app}=0$, numerically solving the three layer coupled LLG equation at T=400K.

### III.B.3. Stability analysis for $j_{app}>0$

The next step is to analyze the attenuation as a function of the applied current to examine how the linear modes are driven into steady state. In the subcritical regime $j_{app}<j_c$ the attenuation of all modes is negative (damped modes). At $j_{app}=j_c$ the attenuation of one of the modes reaches zero, meaning that the linear mode becomes unstable (either transition to another static state or towards a dynamic state). The critical current $j_c$ is therefore defined as the current at which the attenuation reaches zero. We first analyze the range of positive current polarity. In the 'uncoupled picture' this would correspond to reducing the absolute value of the attenuation of the FL excitation and to increasing the attenuation of the ac-SyF mode. The analysis of the current dependence of the attenuation of the coupled modes 1 and 2 is shown in Fig. 2(a-c). At zero current both modes show negative λ (Fig 2a-bottom) as expected from damped modes. When a positive current is applied the absolute value of the attenuation of mode 1 (red line in Fig. 2b bottom) decreases at low fields and the attenuation of mode 2 (green line in Fig. 2b bottom) decreases at high fields. In other words, the attenuation of the modes dominated by the FL decreases in the different field regions. In the intermediate field region around the splitting modes 1 and 2 both have strong contributions from the FL and the SyF. As a consequence the absolute value of the attenuation of these modes changes less upon increasing current. It is also noted that the absolute value of the attenuation of mode 3 increases in the whole field range. Therefore, assuming that instability corresponds to steady state (as has been checked by numerical macrospin simulations) it is found that the steady state oscillations change from mode 1 to mode 2 upon increasing the applied field (Fig 2b top). At intermediate fields both modes remain stable (no steady state excitations occur). If the current is increased further the field range without excitations decreases (Fig. 2c-up). This picture explains well the experimentally observed gaps shown in Figs. 1c and 1d (green symbols).

### III.B.4. Stability analysis for $j_{app}<0$

The same analysis of the attenuation of the coupled modes can be done for negative currents (Fig. 3), where the SyF is predominantly excited. Upon increasing negative current the absolute value of the attenuation of mode 2 is reduced at low fields and the attenuation of mode 1 is reduced at high fields, i.e. in the regions where these modes are dominated by the SyF. Mode 3 is dominated by the SyF in the whole range of field and its attenuation is reduced at all fields, but more strongly at low fields. As a consequence steady state excitations are observed at low fields and at high fields separated by a gap without excitations at intermediate field values (Fig. 3c-top). At high fields, the attenuation of mode 1 is the first one to reach positive values. At low fields, the attenuation of mode 3 is the first one to reach positive values, thus mode 3 is expected to be excited first. This mode is of interest for high frequency applications. Unfortunately mode 3 has too high frequency to be observed within our experimentally accessible frequency range (<20GHz). In the stability analysis it can be seen that for slightly larger negative currents the attenuation of mode 2 reaches also positive values at low fields (see Fig. 3c-bottom). Thus, at low fields and negative currents we expect to have a region with two steady state solutions in competition. Numerical macrospin simulations at T=0 K in this region of field show that the steady state corresponds to mode 3 at low currents and to mode 2 at large currents. This is in good agreement with the experiments, where we observe steady state excitations at frequencies which we attribute to mode 2.

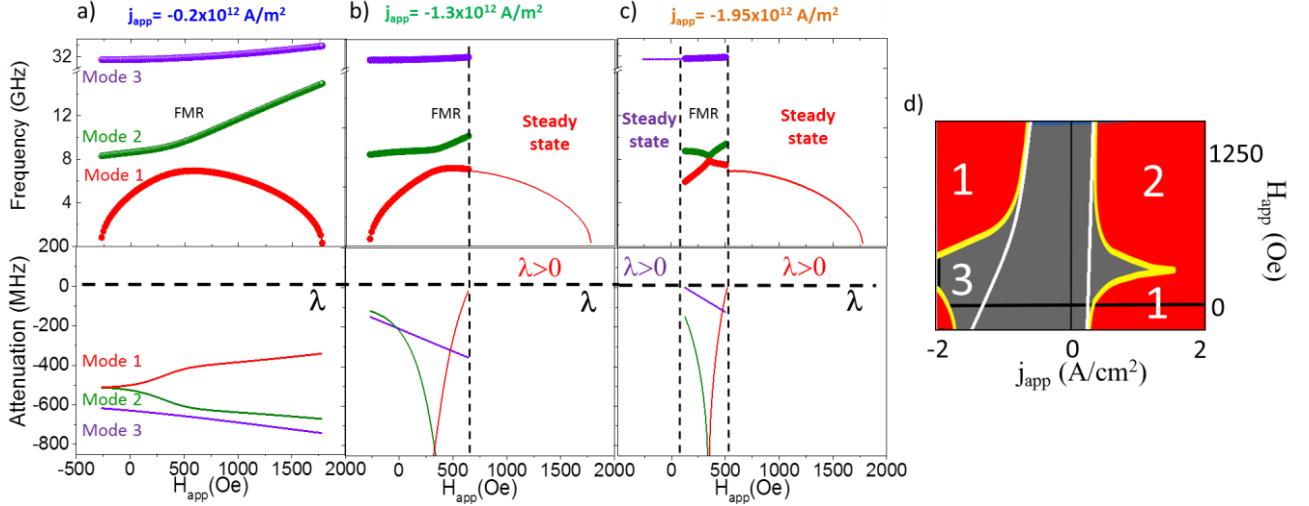

Figure 3: (a-c,top) Frequency-field dependence of the three linear modes of the coupled system (red, green and violet symbols respectively) at different negative currents. Full symbols represent FMR frequencies (i.e. negative attenuation) while continuous lines in c are guides to the eye to highlight the modes that are in steady state (note that the real frequency of the steady state mode will be shifted). (a-c, bottom) Field dependence of the attenuation of the linear modes. (d) Numerical state diagram obtained by linearization of the LLG equations. Grey color represents stable static state and red color represents instability of the static state (numbers indicate which mode is excited in each region). Yellow lines represent critical current of our three coupled layers system and white lines represent critical current when dynamic dipolar field is not taken into account.

### III.B.5. Critical boundaries

The critical currents obtained through this stability analysis give rise to the state diagram shown in Fig. 3d (yellow lines correspond to the critical current), which is in good agreement with the experimental one (Fig. 1b). In particular, both in experiments and simulations the gap of negative current (dominated by the ac-SyF mode) is more pronounced than the one of positive currents (dominated by the FL mode). In order to better understand which interaction is responsible for the gaps of enhanced attenuation, we have performed numerical simulations without dipolar interaction (only RKKY and mutual spin torque) and without mutual spin torque (only RKKY and dipolar interaction). The state diagram when only dipolar interaction and RKKY are considered still exhibits this kind of gaps. On the contrary, white lines in Fig. 3d represent the critical currents obtained when the dipolar interaction is switched off. The gaps without steady state excitations disappear, confirming that the dipolar interaction is the main responsible for this effect [13].

To summarize this section, it is concluded that the gaps observed experimentally in the state diagram and in the frequency-field dispersion (Fig. 1b and 1c), correspond to conditions for which steady state excitations vanish. In this range of field modes 1 and 2 have both equal contributions from the FL and the SyF, which translates into an increased attenuation and thus an increased critical current. This analysis allowed us to identify the character of the steady state modes observed in the experimental state diagram (Fig. 1b) in the four regions of current and field: (i) for low field and positive currents the FL-dominated mode 1 is excited; (ii) for large field and positive currents the FL-dominated mode 2 is excited; (iii) for low field and negative current the SyF-dominated mode 2 is excited; and (iv) for large field and negative current the SyF-dominated mode 1 is excited. The simulations reveal in addition that at low field and low negative current the SyF-dominated mode 3 can be excited, which was not accessible in the experiment.

This analysis of the linear modes and their instabilities leading to different regions of excitations within the state diagram (as compared to non-interacting single layer excitations) is very important to interpret properly experimental observations in spin torque oscillators. For instance, the linear interpolation of the frequency of the FL-dominated steady state mode 1 from low to high fields and mode 2 from high to low field can result in two almost parallel f-H dispersions, that might be wrongly associated to higher order modes of the FL in a non-interacting model, while here we have shown that it actually corresponds to the FL-dominated mode coupled to the ac-SyF mode.

## IV. NON-LINEAR REGIME

In this section we present a detailed analysis of the effect of non-linear interactions that occur when mode 2 is in steady state at high fields and positive currents. Here one can distinguish resonant and non-resonant interactions of the steady state with different damped modes. We start presenting the experimental results and the numerical analysis that goes beyond the one presented in ref. 10.

### IV.A. Experiments

Figure 4a shows the experimental frequency-field dependence. As can be seen for $j_{app}=0.32 \cdot 10^{12} A/m^2$ a discontinuous jump of the frequency occurs at a field of 350 Oe. This jump is accompanied by a hysteresis for which two modes are visible in the spectrum (Fig. 4b). As has been discussed in ref. 10 in this field range there also exists a reduction of the linewidth, which furthermore depends on the RKKY interaction strength within the SyF. Here we present a more detailed analysis of the bistable region with the two modes to better understand the origin of the frequency jump and argue that this is a result of a resonant interaction of the steady state and a damped mode. For this we compare results from numerical simulations realized at T=0K (steady state mode only) and T=400K (steady state and damped modes).

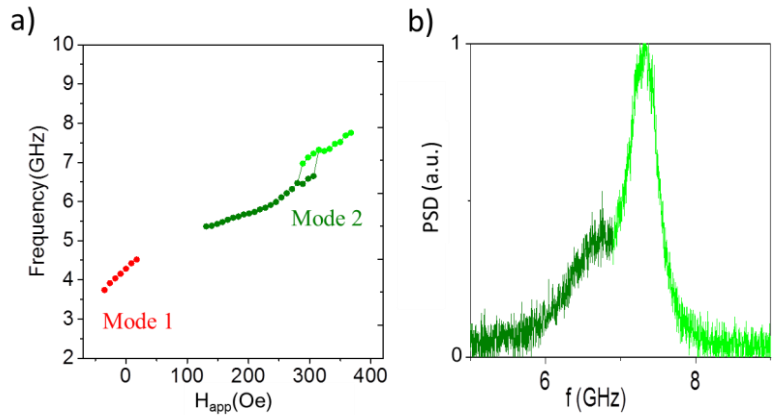

Figure 4: (a) Experimental steady state frequency-field dispersion at $j_{app}= 0.32 \cdot 10^{12}$ A/m$^2$ and (b) experimental spectra at H=300 Oe.

### IV.B. Analysis of the frequency jump

### IV.B.1 Current dependence

Fig. 5 shows simulation results for the evolution of the frequency-field dispersion of the steady state mode 2 upon increasing the applied current, obtained from simulations at 400 K. At low current (pink, black and orange curves), a frequency redshift can be observed upon increasing current. This redshift is much stronger at higher fields (df/dj$_{app}$= 8.4 MHz.µm$^2$/mA at 1560 Oe) than at lower fields close to the splitting (df/dj$_{app}$= 1.5 MHz.µm$^2$/mA at 670 Oe). This is different to a non-interacting free layer excitation that is characterized by a constant frequency redshift for all fields[2, 6]. This different behavior arises from the strong variation of the critical current along the field range (yellow lines in Fig. 2c). j$_c$ is smaller at high fields than around the splitting region (gap). This translates into a higher supercriticality at high fields for same values of applied current.

Upon increasing the applied current (blue, brown and red symbols Fig. 5), the slope df/dH at low fields is much reduced and almost zero, while at larger fields a jump in the frequency is observed, similar to the experiments in Fig. 4a. This jump is accompanied by a hysteresis for which two modes are visible in the spectrum. In the following we will refer to these two modes as the lower and upper frequency branches.

**IV.B.2 Resonant mode interaction**

The discontinuity observed in the f vs H dependence is reminiscent of the mode splitting which takes place when the frequencies of two linear modes cross as described in section III for the FL and SyF modes in the presence of dipolar interaction. This suggests that the frequency jump of Figs. 4, 5 may come from the interaction of the steady state mode 2 with other linear modes of the system. However, Fig. 5 shows that there is no crossing of mode 2 neither with the frequency of the damped mode 3 nor with mode 1. Even when increasing the current and considering the strong frequency redshift of mode 2, its frequencies remain far from mode 1. The steady state mode 2 is an in-plane precession mode that is characterized by emission not only at its fundamental frequency, but also at higher harmonics, in particular its third harmonics. In order to investigate the possibility of mode interaction via the third harmonics of mode 2 we plot the frequencies of mode 3 for zero current at one third of its values in Figs. 5, 6a and 6b (violet continuous line). Star symbols in Fig. 6b correspond to one third of the frequency of mode 3 in presence of a positive current. The frequencies are little affected by the current. From this it can be seen that the frequencies of the 3$^{rd}$ harmonics of mode 2 are close to the linear mode 3 in the region where the hysteresis exists. To better elucidate whether there is a true mode interaction, we show in Fig. 6c the FFT spectra of the M$_y$-component of the FL magnetization, in the frequency ranges around mode 2 and around mode 3 as a function of the applied field at constant current. Spectra on the left correspond to the lower frequency range, around mode 2. Spectra on the right correspond to higher frequency range around 30 GHz, around the third harmonic of mode 2 and the damped mode 3. Let us analyze the figure going from lower fields to higher fields (i.e. from Fig. 6c top to Fig. 6c bottom).

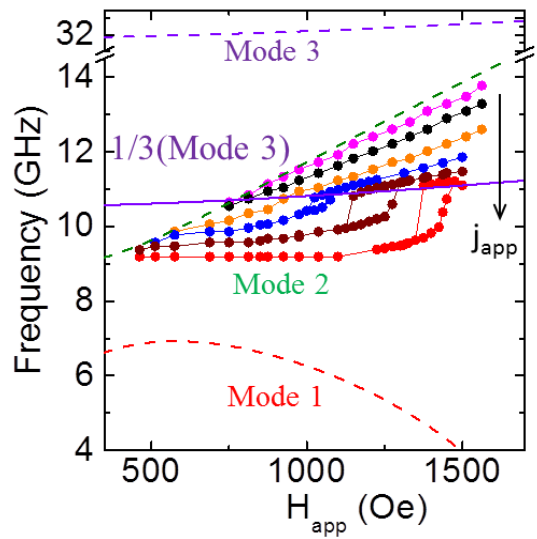

Figure 5: Evolution of the simulated frequency field dispersion of the steady state at 400 K upon increasing current, for j$_{app}$= 0.3, 0.4, 0.6, 0.7, 0.8 and 0.9 x 10$^{12}$A/m$^2$ (pink, black, orange, blue, brown and red symbols respectively).

At H=1300 Oe, the frequency of the steady state mode 2 is around 9.5 GHz and its third harmonic is around 28.5 GHz, while the frequency of mode 3 is at 33 GHz. As expected, we observe only one excitation peak in the low frequency range and two peaks at the higher frequency range. Upon increasing field, the frequency of mode 2 shifts towards mode 3 whose frequency does not change much. At H=1400 Oe these modes interact and an additional peak appears at low frequency (around 11 GHz, one third of the frequency of mode 3), i.e. two well defined peaks clearly coexist in the higher *and* lower frequency range (see Fig 6c, H=1400 - 1440 Oe). In contrast to the two peaks at higher frequency, the appearance of two peaks at lower frequency is unexpected and would not occur in absence of interaction. It can qualitatively be explained through a non-linear interaction of mode 3 with mode 2 through its third harmonic. When mode 2, that is FL dominated, is in steady state its amplitude increases and with this the associated dynamic dipolar field. This dipolar field is sensed by the SyF dominated mode 3 which can be pumped in a resonant manner when the frequency of the dipolar pumping field is close to its resonance frequency. This is the case when the frequency of the 3$^{rd}$ harmonics of mode 2 is close to the one of mode 3. As a consequence mode 3 is driven into resonance and due to its increased amplitudes starts to strongly interact with mode 2. This mode hybridization increases the intensity of the higher frequency mode while the one of the lower frequency mode decreases, until it is more or less absent at fields of H=1500 Oe. As discussed in ref. 10, the linewidth of this strongly hybridized peak is much reduced.

This analysis of the FFT spectra along lower and upper frequency branches clearly shows that the steady state mode 2 interacts in a resonant manner with the damped mode 3, leading in a certain field range to a new hybridized mode (upper branch). The fact that it is the frequency of the steady state mode 2 that locks onto the one of the damped mode 3 can be qualitatively explained through its higher agility which results from the non-linear dependence of its frequency on amplitude. Thus the FL-dominated steady state mode 2 can adjust its amplitude and with this its frequency more easily than the linear SyF-dominated damped mode 3. This is illustrated in Fig. 6 that shows the FL and SyF layer trajectories of the upper (Fig. 6e) and lower (Fig. 6d) frequency branch for T=0 K at a field of H=1250 Oe and two values of current. The FL shows a reduction in the amplitude that leads (according to its redshift character) to an increase in frequency when comparing the lower to the upper branch. This change of amplitude of the FL magnetization is accompanied by more complex trajectories of the SyF-layers.

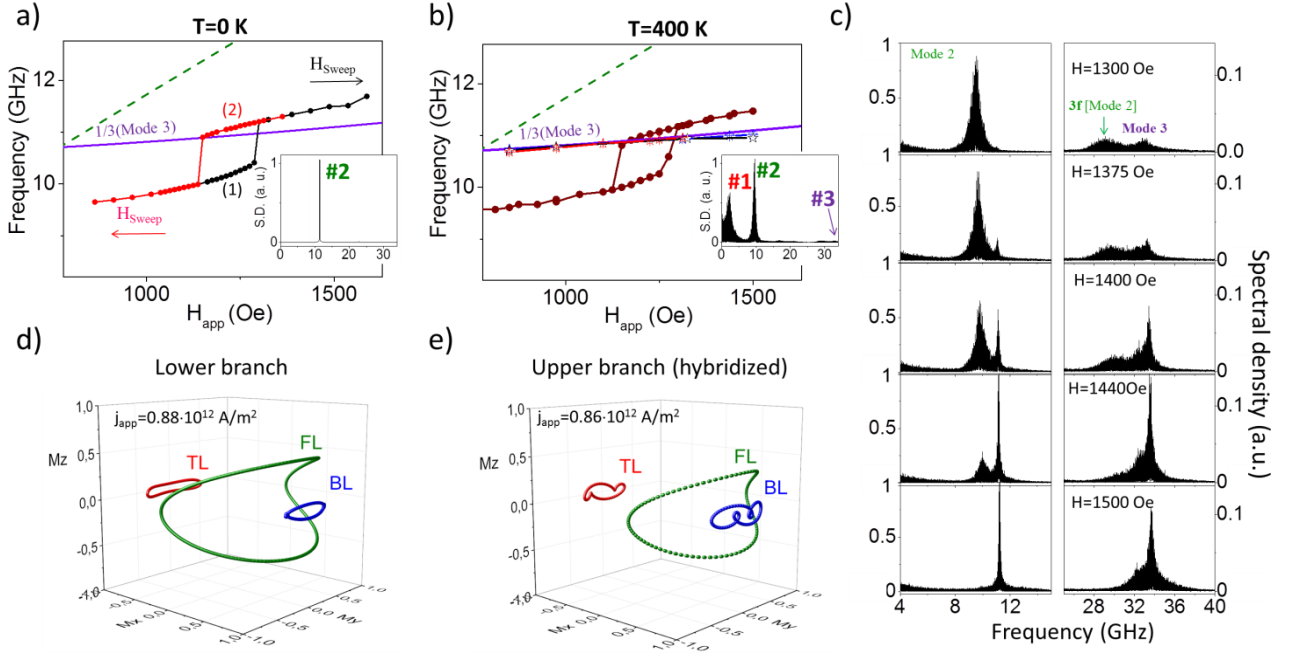

Figure 6: (a-b) Detail of the frequency behavior around the kink/jump effect for $j_{app}$= 0.8 x 10$^{12}$A/m$^2$, obtained (a) at T=0K in both sweeping directions of the applied field and (b) at T=400 K. (c) Simulated FFT spectra obtained at $j_{app}$= 0.9 x 10$^{12}$A/m$^2$ at different fields in the low (left) and high (right) frequency range. (d-e) Variation of the trajectories at a constant field of H=1250 Oe upon increasing current, just before and after the jump: (d) corresponds to the lower frequency branch ($j_{app}= 0.88 \cdot 10^{12}$ A/m$^2$), and (e) to the upper frequency branch ($j_{app}= 0.86 \cdot 10^{12}$ A/m$^2$).

**Hysteresis and coexistence of two modes**

The next question to address is whether the two modes in the hysteretic region in Fig. 6c are both steady state modes and how they do coexist. In order to investigate this, we compare the simulation results obtained at T=400K with T=0K simulations for different sweeping direction of the field. Note that simulations at T=0 K provide only solutions for steady state modes (see Fig. 6a-inset) but not on damped modes. The results are shown in Fig. 6a for T=0K. The hysteretic part is a true hysteresis and depends on the field sweeping direction. Going from low to high fields only the lower frequency branch exists (black points) that then jumps abruptly to the higher frequency branch. Going from high to low fields, only the higher frequency branch exists, that jumps abruptly to the lower frequency branch (red points). From this it can be concluded that both branches are steady state solutions, where the lower branch corresponds to the steady state mode 2 and upper branch to the steady state mode 2 strongly hybridized with mode 3.

When temperature is included the modes of both branches appear in the spectra, see Fig. 6b,c. Using the frequency-time spectrogram analysis it is found though that the system hops in time between the two branches. This is shown in Fig. 7, obtained from the $M_y$–component of the FL at T=400 K. Hence, while both modes are steady solutions, they do not coexist in time but hop through thermal activation from one to the other. It is noted that a similar situation has been discussed theoretically in [19] for two steady state solutions that can exist within a single layer. It is shown that thermal excitations lead to non-zero amplitude of both modes.

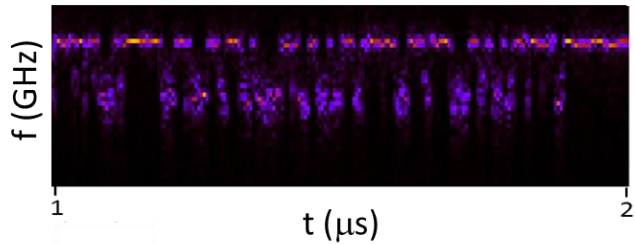

Figure 7: Frequency-time spectrogram obtained from the $M_y$ of the FL at T=400K.

**Non-linear interaction**

As a final comment we want to stress that the resonant interaction is a non-linear interaction effect. As already mentioned, mode 3 is a damped mode whose attenuation increases with positive current (i.e. it becomes more damped and is not a steady state mode). The discussed effect thus arises from the interaction of the steady state mode 2 with a truly linear mode 3 of the system. By switching on and off the different dynamic interactions of the coupled system in the simulations, we have identified the dynamic dipolar interlayer interaction as the main responsible interaction. The mutual spin transfer torque is found to play a minor role, which in our case translates only into a small reduction of the current required to observe the effect. The fact that it is an effect mediated by dipolar interaction may seem reminiscent of the mode splitting occurring when two linear modes interact through dipolar coupling [17]. However, it should be pointed out that the jump in the frequency of the steady state mode 2 due to interaction with mode 3 is not observed when the two modes are linear modes or when the precession amplitude of the steady state mode 2 is low (at lower current, see Fig. 5). It only occurs at higher currents and higher amplitudes. This indicates that it is essentially a non-linear resonant interaction of the steady state with a damped mode. From this it can also be concluded that there is a difference between two modes that are first coupled in the linear regime and then driven into steady state (see section III), and a mode that is first driven into steady state and then interacts with another linear mode.

**IV.B.3 Non-resonant mode interaction: Frequency redshift of mode 1**

The simulations also reveal the effect of the steady state mode 2 on the damped mode 1, even though there is no frequency crossing and hence no resonant interaction. Mode 1 is a linear mode which is more damped the more positive current is applied, see Fig. 2. From T=400K simulations we have extracted its frequency vs field for different positive current values, as shown in Fig. 8. Two interesting features are observed:

(1) Mode 1 shows a very strong frequency redshift with increasing positive current, for instance the frequency is down shifted by 3 - 4 GHz as compared to the zero current frequency (red curve in Fig. 8) for a current density of $j_{app}= 0.9 \cdot 10^{12}$ A/m$^2$ (brown curve in Fig.8). This behavior differs strongly from the typical trend of non-interacting single layer linear modes, for which a small frequency variation is expected upon changing the current due to an increase of effective damping produced by the positive current. Qualitatively, we explain the frequency change of mode 1 as a non-resonant interaction with the steady state mode 2. Due to the large amplitude of the steady state excitation of mode 2, the damped mode sees a modified time averaged dipolar field arising from mode 2. This results in a change of the frequency of mode 1 and happens within the whole range of field. Another important consequence of this interaction is a reduction of the spin flop field under increasing current.

(2) In the hysteretic region, the frequency of mode 2 changes non-continuously, see Figs. 5, 6 and with this also its precession amplitude and dipolar field (see Fig. 6d,e). This modified amplitude is also reflected in the frequency of mode 1 that shows a flattening (see Fig. 8) around the hysteretic region of mode 2, for all three current values.

These two effects on mode 1 can only be explained as manifestations of the strong impact that a non-linear mode has on the damped modes of the system, even if they do not interact directly (i.e. it is a non-resonant interaction). More generally it can be concluded that any changes in one of the modes will have consequences for the other modes. This is a general important consequence of the non-linear coupling within the system. In the following discussion a qualitative model is used to support the effect of resonant and non-resonant mode interactions in the non-linear regime.

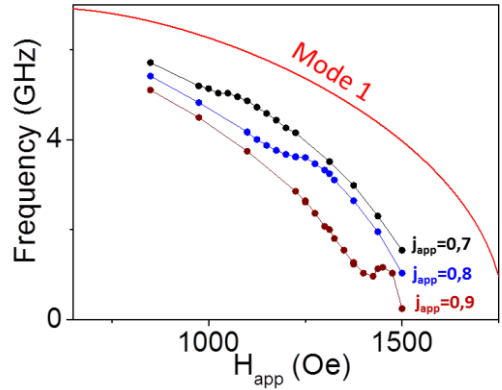

Figure 8: Frequency-field dispersion of the damped mode 1 at $j_{app}= 0 \cdot 10^{12}$ A/m$^2$ (red line) and under application of positive current $j_{app}= 0.7, 0.8$ and $0.9 \cdot 10^{12}$ A/m$^2$ (color dot symbols).

**IV.B.4 Discussion**

In the following we would like to present the results of section IV in a more general picture using recent results from an analytic description of the steady state excitation of a more simple model system of two layers coupled via conservative exchange interaction and spin momentum transfer [14]. Upon replacing one of the layers by a SyF structure, and the conservative exchange by the conservative dipolar interaction one can make a direct analogy to the three layer model system discussed here.

From the analytic description in ref 14, it was shown that when one mode of the system is driven into large amplitude steady state, the linear eigenmodes of the coupled system do not remain eigenmodes in the non-linear regime. Instead, the resulting steady state mode as well as the damped modes are hybridizations with contributions from all the linear eigenmodes through conservative or dissipative coupling terms. In the analytic description of the two layer model system this can be clearly seen when writing down the equations of motions for the complex amplitudes of the two eigenmodes and the corresponding power and phase equations (see appendix). Here we are discussing effects arising mainly from conservative coupling (dipolar interaction), i. e., from the coupling terms of the Hamiltonian. The conservative coupling makes contributions to the phase and power equations of the hybridized modes through terms that can be divided in:

(i) Parametric interaction between modes[20, 21]: Non-linear coupling terms of higher order proportional to the power (amplitude) but which include also phase relations modes (F and S terms in equation 4 of Ref. 14 and equation 1 in the appendix). They allow energy transfer between modes. These terms depend strongly on the frequency difference and may become resonant when the frequencies of two modes (or their harmonics) are close.

(ii) Non-linear frequency shifts: non-linear coupling terms proportional only to the power and coming from the terms of order 2 of the Hamiltonian (T term in equation 4 Ref. 14 and equation 1 in the appendix). These terms give rise to non-resonant interactions, and become important when one mode is in steady state (large amplitude).

Within this analytical model, the interaction discussed in section IV.B.2 of the steady state mode 2 with the damped mode 3, is identified as a parametric resonant interaction. Since the amplitude of the steady state mode 2 is large, the non-linear F and S coupling terms can become large and non-zero (i.e. resonant) when the frequencies (or their harmonics) of the corresponding modes are close. These non-zero contributions to the frequency and power equations will lead to a strong hybridization of the two modes and eventually results in the frequency jump of the steady state mode 2 (Figs. 4, 5 and 6), whose frequency is more drastically shifted than mode 3 because of its higher tuning capabilities. We note that a similar interaction of steady state and damped modes via their harmonics has been reported in Ref. 18 for vortex excitations.

Finally, as described in section IV.B.3, mode 2 being in steady state (i. e. having large amplitude) translates into a modification of the averaged dipolar field felt by mode 1. Within the analytical model this is identified as a non-resonant interaction of mode 2 with mode 1 through the non-linear frequency shift (T terms) that leads to a strong frequency redshift of mode 1 along the whole field range. This redshift is more pronounced at larger fields than at lower fields (see Fig. 8). The reason is that within this range of field the supercriticality increases with field at a constant current (see Fig. 2c-bottom). Therefore, at a constant value of current, the larger the field, the larger the oscillation amplitude of the steady state. In principle, this effect produced by mode 2 being in steady state should apply as well to mode 3, which nevertheless does not show a strong frequency shift. To understand this it is worth having another look at Fig. 2d(3)-(6). As can be seen in the figure, in this region of field mode 3 is an *almost pure* SyF mode, with a negligible contribution from the free layer (Fig.2d(3)) and thus from (the free layer dominant) mode 2 via the T terms. This explains why mode 2 being in steady state has a smaller impact on this mode 3 than on mode 1, which has a larger contribution from the free layer.

## V. SUMMARY

To summarize, in this work we have shown that dynamic interactions are needed to understand the spin torque driven excitation spectra of a standard STO. In particular, the dynamic dipolar coupling plays a major role. It leads to a hybridization of the linear modes of the free layer and the SyF, characterized by a mode splitting. It is responsible for the field region of enhanced attenuation which translates into gaps in the frequency-field dispersion of the steady state. Finally, it is responsible for the resonant and non-resonant non-linear interactions between the spin torque driven and other damped modes. As was shown here, resonant interactions translate into discontinuous deviations or jumps in the steady state frequency-field and frequency-current dispersions. When this happens, the steady state evolves into a two mode regime characterized by thermally activated mode hopping. Non-resonant interactions lead to a strong frequency redshift of the lower frequency damped mode along a wide range of field. These results underline the strong impact of coupling in the excitation spectra of spin torque oscillators and that in a coupled system the steady state and damped modes can influence each other.

We acknowledge B. Ocker, J. Langer, and W. Maas from Singulus Technologies for material deposition of the spin valve structures. This work was supported in part by the French National Research Agency (ANR) under Contract No. 2011 Nano 016-07 (SPINNOVA) and in part by the EC under the FP7 program No. 317950 MOSAIC.

**Appendix 1:**

After applying two canonical transformations [12] the Hamitonian of the model two layers system reads:

$$\mathcal{H} = \omega_1 |b_1|^2 + \omega_2 |b_2|^2 + \frac{1}{2} N_1 |b_1|^4 + \frac{1}{2} N_2 |b_2|^4 + T |b_1|^2 |b_2|^2$$
$$+ \frac{1}{2}\left(F b_1 b_1 b_2 b_2 + S b_1 b_1 b_2^* b_2^* + c.c.\right) + H_{4r} + H_{hr} \quad (1)$$

Where the coefficients, $\omega_1$, $\omega_2$, $N_1$, $N_2$, $T$, $F$, $S$, are real and depend on the material parameters related to $H_{eff,i}$ and to the RKKY coupling fields $H_{RKKY,i}$, but are independent of the current and of the Gilbert damping constant.

From this Hamiltonian the conservative (C) part of the equation of motion is obtained:

$$\dot{b}_i\Big|_C = -j \frac{\partial \mathcal{H}}{\partial b_i^*} = -j(\Omega_i + \Psi_i) b_i$$

using the following definitions:

$$\Omega_1 = \omega_1 + N_1 p_1 + T p_2$$
$$\Psi_1 = F \frac{b_1^* b_2^{*2}}{b_1} + S \frac{b_1^* b_2^2}{b_1} + \frac{1}{b_1} \frac{\partial (H_{4r} + H_{hr})}{\partial b_1^*}$$
$$\Omega_2 = \omega_2 + N_2 p_2 + T p_1 \quad (2)$$
$$\Psi_2 = F \frac{b_2^* b_1^{*2}}{b_2} + S \frac{b_2^* b_1^2}{b_2} + \frac{1}{b_2} \frac{\partial (H_{4r} + H_{hr})}{\partial b_2^*}$$

where the powers $p_i$ are defined as $p_i = |b_i|^2$, $\Omega_{1,2}$ are real and $\Psi_{1,2}$ are complex through its dependence on $b_1$, $b_2$, $b_1^*$ and $b_2^*$ while $F$ and $S$ are real numbers.

Finally, by applying the transformation also to the dissipative terms of the LLG equations (see ref. 12), and considering solutions to the general perturbed Hamiltonian equations of the form $b_i(t) = |b_i(t)| \exp[-j\phi_i(t)]$ with $i=1,2$, two equations for the amplitudes and two equations for the phases are obtained:

$$\begin{cases} \dot{p}_1 = -2[\text{Re}(\Gamma_1) + \text{Im}(\Psi_1) + \text{Im}(\Sigma_1) + \text{Re}(\Pi_1)] p_1 \\ \dot{p}_2 = -2[\text{Re}(\Gamma_2) + \text{Im}(\Psi_2) + \text{Im}(\Sigma_2) + \text{Re}(\Pi_2)] p_2 \end{cases}$$

(3)

$$\begin{cases} \dot{\phi}_1 = \Omega_1 + \text{Re}(\Psi_1) + \text{Re}(\Sigma_1) + \text{Im}(\Gamma_1) + \text{Im}(\Pi_1) \\ \dot{\phi}_2 = \Omega_2 + \text{Re}(\Psi_2) + \text{Re}(\Sigma_2) + \text{Im}(\Gamma_2) + \text{Im}(\Pi_2) \end{cases}$$

Where $\Sigma_i$, $\Gamma_i$, $\Pi_i$ are complex functions coming from the dissipative part of the LLG equation.